\documentstyle[epsfig]{mn}
\def\etal{{\it et al}}
\def\deg{^{\circ}}
\def\P3hat{{\mathaccent 94 P}_3}

\def\eg{{\it e.g.}}

\newcount\notenumber
\def\clearnotenumber{\notenumber=0}
\def\note{\advance\notenumber by1 \footnote{$^{\the\notenumber}$}}
\clearnotenumber
\begin{document}
\title[`Notches' in the Profiles of Bright Pulsars]
	{`Notches' in the Average Profiles of Bright Pulsars}
\author[McLaughlin \& Rankin]
{Maura A.\ McLaughlin$^{1}$ \& Joanna M.\ Rankin$^{2}$\\
$^1$Jodrell Bank Observatory, University of Manchester, Macclesfield UK SK11 9DL: mclaughl@jb.man.ac.uk \\
$^2$Physics Department, University of Vermont, Burlington, VT 05405 USA : joanna.rankin@uvm.edu}

\date{}
\maketitle

\begin{abstract}
We discuss the discovery of `notch-like' features in the mean pulse 
profile of the nearby, bright pulsar B0950+08.  We compare these 
low-level features with those previously seen in the pulse profiles 
of pulsars J0437$-$4715 and B1929+10. While J0437$-$4715 
is a binary millisecond pulsar and B0950+08 and B1929+10 
are isolated, slow pulsars, all three pulsars are nearby and very bright. 
Furthermore, all three have detectable emission over an unusually wide 
range of pulse phase.  We describe the similar properties of the notch 
features seen in all three pulsars and discuss possible interpretations.

\end{abstract}

\section{Introduction}

Detailed studies of the structure of radio pulsar pulse profiles are 
essential for understanding the pulsar emission mechanism and 
the forms of the pulsar emission beam. Pulse profiles are remarkably 
stable over timescales ranging from a few pulses to many years, 
indicating that these profiles are determined by permanent physical
characteristics of the neutron star and its environment. Many profile 
properties can be explained very well by the ``hollow-cone'' model 
of pulsar emission developed by Radhakrishnan \& Cooke (1969) 
and Komesaroff (1970), shortly after the discovery of the first pulsar. 
This ``magnetic-pole'' model, further developed by Sturrock (1971) 
and Ruderman \& Sutherland (1975), involves charges streaming 
from the polar cap along the open field lines, radiating in the direction 
of their motion and forming an emission cone aligned with the magnetic 
axis of the star.  As more pulsars were discovered, many were found 
to have two, three or even more pulse components, prompting various 
efforts to classify and interpret the various forms (\eg. Backer 1973; 
Rankin 1983, 1993; Lyne \& Manchester 1988). 

In this paper we draw attention to the strange ``notches'' in the profiles 
of B0950+08 and two other pulsars, J0437--4715 and B1929+10.  All 
three stars exhibit very similar double ``notch" features which appear 
to have virtually nothing to do with the component structures of their 
very different profiles.  B0950+08 and B1929+10 have rotational periods 
of 253 and 227 ms, whereas J0437--4715 is a binary millisecond pulsar 
with a period of only 5.75 ms.  Significantly, these stars are some of the 
very closest among the pulsar population---all with dispersion measures 
(DMs) of about 3 ~pc cm$^{-3}$ and parallax-determined distances of some 
262$\pm$5, 331$\pm$10 and 170$\pm$3 pc (Brisken \etal\ 2000, 2002)---and, 
in spite of the fact that they all have rather ordinary spindown values, all 
have been detected in the optical and/or soft x-ray region (Mignani \etal\ 
2002; De Luca \etal\ 2003; Wang \etal\ 1997; Zavlin \etal\ 2002).  
B0950+08 and B1929+10 are among the brightest pulsars in the radio 
sky\cite{ppcat}, and J0437--4715 is well known as the brightest 
millisecond pulsar.   In \S\ref{sec:obsres} we discuss our observations of 
B0950+08, present its average pulse profiles and describe the unusual
features.  In \S\ref{sec:discuss} we compare these features with similar
ones seen in the profiles of J0437$-$4715 and B1929+10 and 
discuss possible physical interpretations.

\section{Observations and Results} \label{sec:obsres}

Observations presented in this section were made using the 305-m Arecibo 
Telescope in Puerto Rico.  Observations at 430~MHz were obtained using 
the Arecibo Observatory Fourier Transform Machine 
(AOFTM\footnote{\verb+http://www.naic.edu/~aoftm+}) with 1024 channels across 
a 10-MHz bandpass and with 409.6-$\mu$s sampling. B0950+08 was 
observed at 430~MHz on MJDs 51182, 51188 and 51189 (4th, 10th and 
11th of January, 1999) for durations of 3000, 3700 and 2500~seconds, 
respectively.  The 1475-MHz data presented were acquired with the 
Wideband Arecibo Pulsar Processor (WAPP\footnote{\verb+http://www.naic.edu/~wapp+}) 
with 128 channels across a 100-MHz bandpass and 256-$\mu$s sampling.
Observations of B0950+08 at this frequency were taken on MJDs 52187 
and 52189 (5th and 7th of October, 2001) for durations of 1100 and 7200 
seconds, respectively.

\begin{figure}
\begin{center}
\epsfig{file=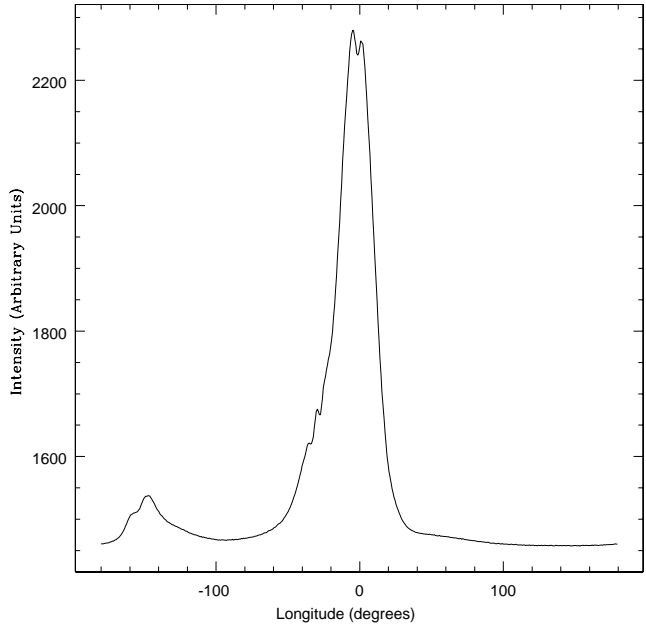, height=9.0cm}
\caption{Pulse profile for B0950+08 taken on MJD 51188 at 430~MHz 
for a total observation length of 3700~s.  A bandwidth of 10~MHz 
and sampling time of 409.6~$\mu$s were used. Data were folded 
using 512 bins across the pulse period of 253~ms.  The ``notch''  
pair at a longitude of $-30^\circ$ is apparent. 
\label{fig:0950.51188}}
\end{center}
\end{figure}

\begin{figure}
\begin{center}
\epsfig{file=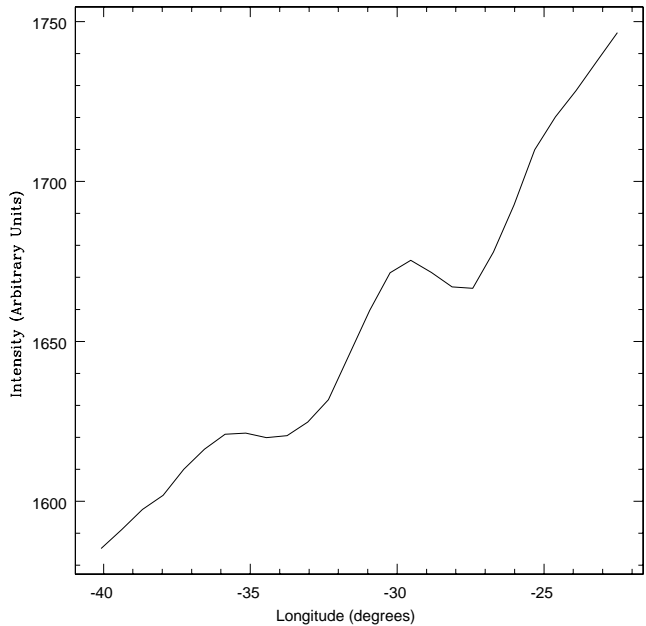, height=9.0cm}
\caption{The profile of Fig.~\ref{fig:0950.51188} with a smaller 
longitude scale.  The notch feature at longitude  $-30^\circ$ is well 
defined. Identical features with the same morphology and phase 
were seen at this frequency in observations on MJDs 51182 and 
51189.
\label{fig:0950.51188.2}}
\end{center}
\end{figure}

\begin{figure}
\begin{center}
\epsfig{file=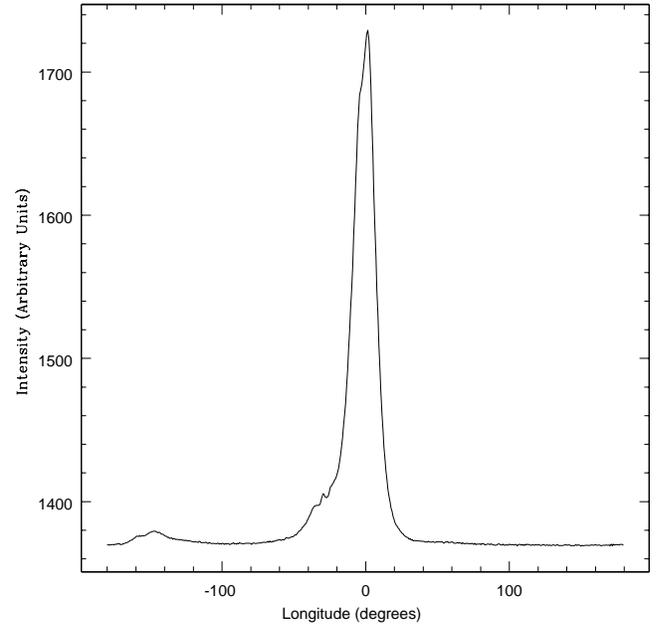, height=9.0cm}
\caption{Pulse profile for B0950+08 taken on MJD 51189 at 430~MHz for a total 
observation length of 2500~s.  Note that while this profile
and that of Fig.~\ref{fig:0950.51188} 
have very different forms, the ``notch''  pair 
at $-30^\circ$ falls at precisely the same longitude (see text). 
\label{fig:0950.51189}}
\end{center}
\end{figure}

\begin{figure}
\begin{center}
\epsfig{file=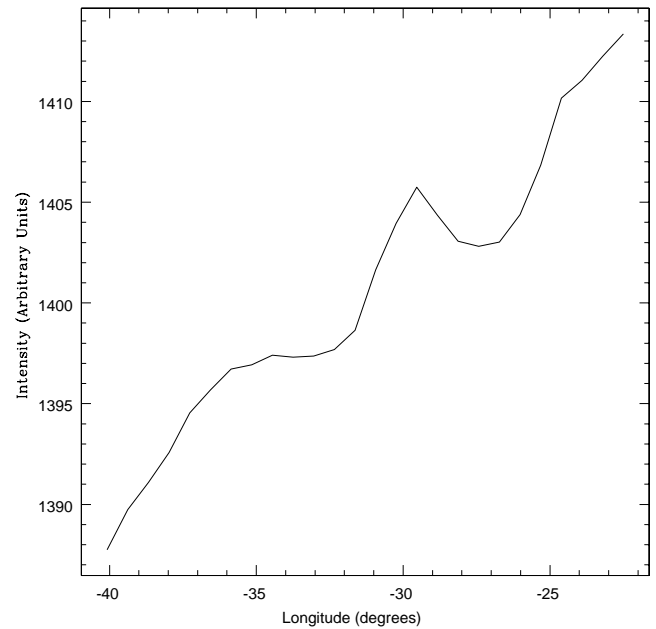, height=9.0cm}
\caption{The profile of Fig.~\ref{fig:0950.51189} with a smaller longitude scale. 
Here again, the feature at $-30^\circ$ is clear. 
\label{fig:0950.51189.2}}
\end{center}
\end{figure}

\begin{figure}
\begin{center}
\epsfig{file=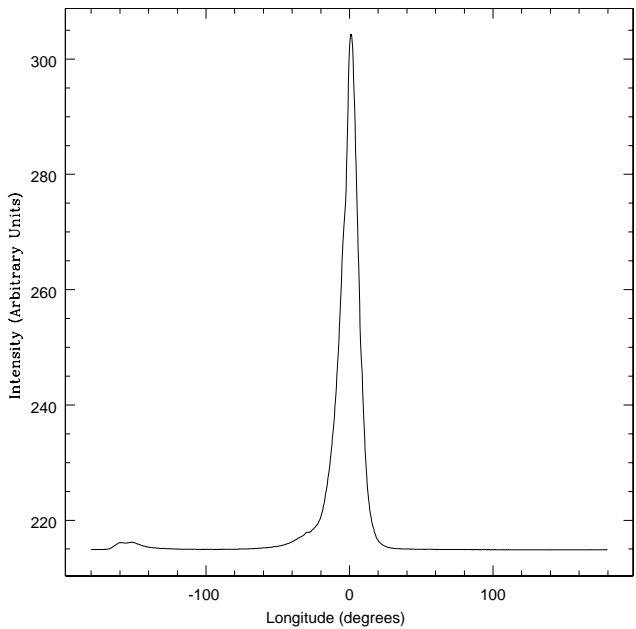, height=9.0cm}
\caption{Pulse profile for B0950+08 on MJD 52189 at 1475~MHz for a total 
observation length of 7200~s. A bandwidth of 100~MHz and sampling 
time of 256~$\mu$s were used. Data were folded using 1024 bins 
across the pulse period of 253 ms.  The notch feature at longitude 
$-30^\circ$ is barely apparent on this scale.
\label{fig:0950.52189}}
\end{center}
\end{figure}

\begin{figure}
\begin{center}
\epsfig{file=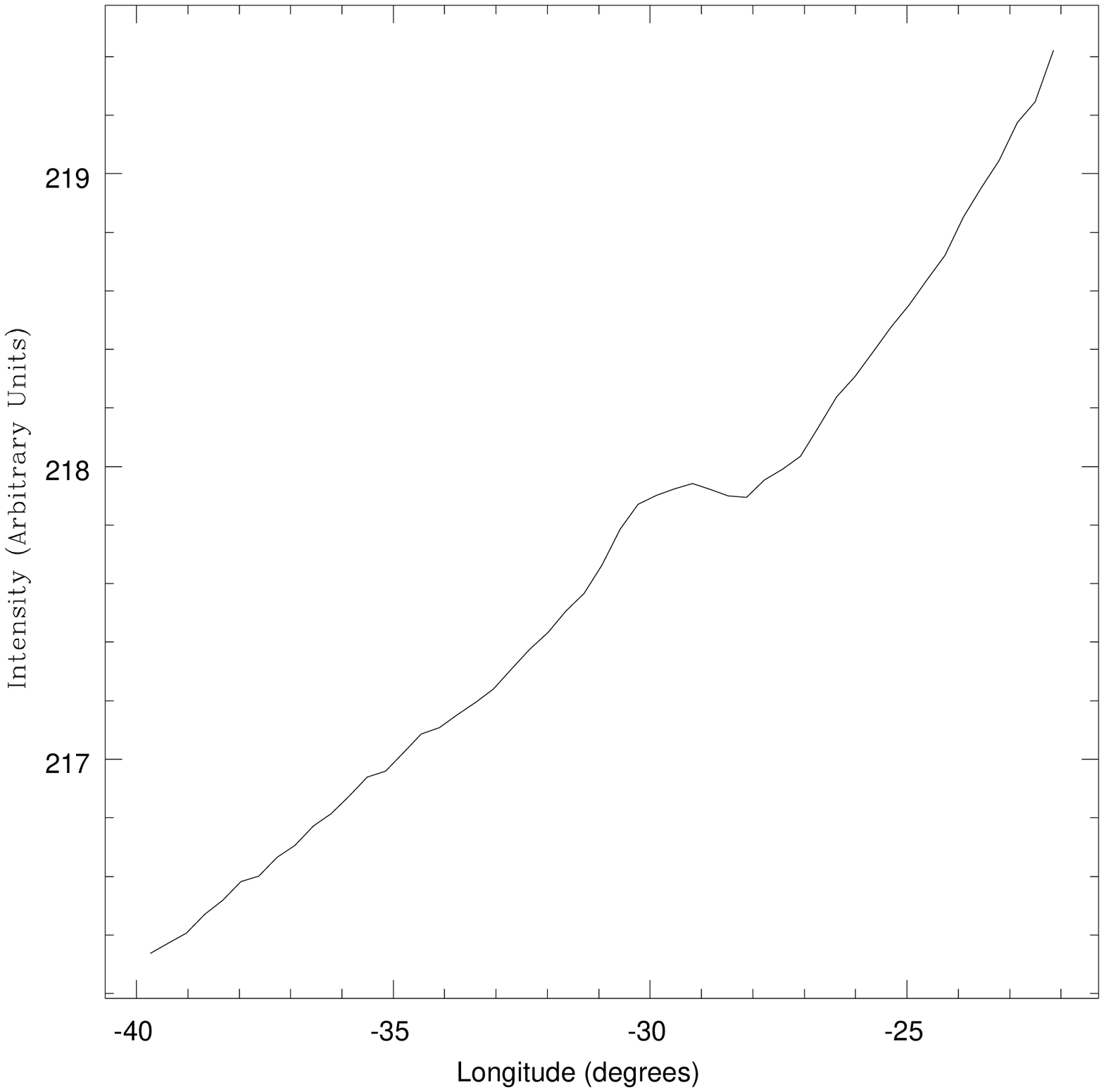, height=9.0cm}
\caption{The profile of Fig.~\ref{fig:0950.52189} with a smaller longitude scale. 
Here the feature at $-30^\circ$ is clear. An identical feature with the same 
morphology and phase was seen on MJD 52187.
\label{fig:0950.52189.2}}
\end{center}
\end{figure}

For each B0950+08 observation, pulse phases were calculated using the 
software package TEMPO\footnote{\verb+http://pulsar.princeton.edu/tempo+} \cite{tay89}  
and an ephemeris obtained at Jodrell Bank Observatory \cite{hobbs2002}.
Data were folded to create mean pulse profiles. The up-to-date ephemeris
allowed alignment of the profiles acquired on different dates and at different 
frequencies.  The 430-MHz average profiles for the MJD~51188 and 51189 
observations are presented in Figs.~\ref{fig:0950.51188} and \ref{fig:0950.51189}.  
Note that the two profiles are distinctly different in overall form, but that the 
low-level features are seen at just the same $-30^\circ$ longitude before the 
main pulse (MP) peak (see Figs.~\ref{fig:0950.51188.2} \& \ref{fig:0950.51189.2}).  
Moreover, a similar pair of features is seen in the MJD~51182 data (not shown).  

In Fig.~\ref{fig:0950.52189} we present a corresponding 1475-MHz 
profile for the MJD~52189 observation, where again a low-level feature is 
seen at some $-30^\circ$ (see Fig.~\ref{fig:0950.52189.2}).  At 430~MHz 
the feature consists of two dips in the profile intensity with half widths of some 
2$\deg$ and their centers separated by some $7^\circ$.  The total width of 
the feature at 430~MHz is $11^\circ$.  At 1475~MHz---in both this observation 
and another (not shown) on MJD 52187---the feature appears as a single 
peak with a width of $3^\circ$. 

Several published observations also appear to record the B0950+08 
``notches''.  They are very clear in the 430-MHz time-aligned profile of 
Hankins \& Rankin (2003), but cannot be discerned in the lower quality 
observations at various other frequencies. Moreover, the ``notches'' can 
probably be seen in the polarimetric observations of both Gould \& Lyne 
(1998) and Weisberg \etal\ (1998).  The former very high quality profiles 
seem to show pairs of features at the correct phase at 408, 610 and 925 
MHz, and in the latter we appear to see a {\it pair} of features at 1418 
MHz.  This latter observation (see their Fig.~5) is especially interesting 
because the features appear to involve the linear polarization also, 
which is only 20--30\% in the region on the far leading edge of the profile.  

It must be said that the features in B0950+08 are rather weak, at best 
reducing the intensity at their centers by a few percent.  It might even 
be regarded as surprising that such features would be found in a star 
which is well known for its sporadic emission (It may not be known, for 
instance, whether the pulsar ever nulls, because the range of intensities 
exhibited by its pulses is so large.)  In this context, it is interesting that 
Nowakowski \etal\ (2003) have shown that this pulsar's profile is 
comprised of different intensity fractions which have quite different 
partial-average profile forms.  These studies indicate that the MP has 
three main ``components'': the trailing one which is usually strongest, 
the middle one which varies in relative intensity, and a leading one 
which can only be discerned in certain populations of weak pulses.  It 
would therefore appear that the distinct 430-MHz profiles seen above 
represent distinct profile ``modes'' in this star.  Finally, the above study 
finds that the ``notches'' are seen only in the very weakest populations 
of pulses, which nonetheless contribute significant power to the profile.  
The authors believe that the weakest pulses represent emission from 
rather higher altitudes than the stronger fractions.

\section{Discussion} \label{sec:discuss}

\begin{figure}
\begin{center}
\epsfig{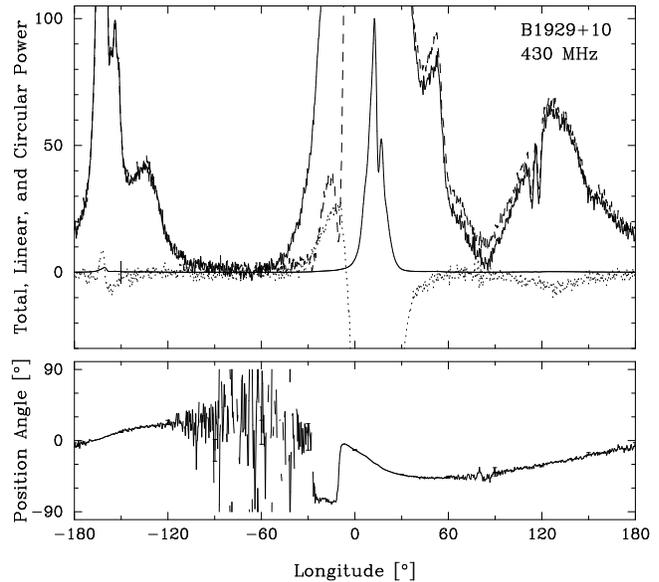}
\caption{Full-period average polarization profile of B1929+10 at 430 MHz 
after Rankin \& Rathnasree (1997).  The total power (Stokes $I$) is plotted 
first at full scale and then at a $\times$250 expanded scale, so that only 
features below 0.4 percent of the MP amplitude are now visible.  The linear 
($L$) and circular ($V$) polarization are then plotted at the expanded scale.  
The PA is plotted in the lower panel.  Note the double ``notch'' feature which 
follows the MP by about 100$\deg$.  This region exhibits complete linear 
polarization (see text), so the ``notches'' are clearly visible in both $I$ and 
$L$, representing an intensity decrease at their centers of some 40\%.  
\label{fig:1929}}
\end{center}
\end{figure}

Unusual features similar to those presented in \S\ref{sec:obsres} have 
been observed in two other pulsars. Rankin \& Rathnasree (1997) reported 
a double notch-like feature in the average profile of B1929+10 at 
430~MHz, and a display similar to theirs using the same observations is 
given in Fig.~\ref{fig:1929}.  As can be seen there, the ``notches'' 
follow the main pulse peak by roughly $100^\circ$ and have a total width 
of approximately $10^\circ$.  More accurate scaling reveals that they have 
widths of some 2.3$\deg$ and that their centers are separated by just twice 
this interval.  Note also that B1929+10's profile is generally fully linearly 
polarized, so that the features are most clearly seen in both the total power 
and total linear polarization\footnote{The possible origin of this strange 
``superpolarization'', that was also seen by Phillips (1990), is discussed in 
Rankin \& Rathnasree (1997).  Phillips, however, seems to have plotted 
his Figs.~1 \& 2 with some unreported smoothing, so that the ``notches'' are 
not visible, though there is a broad dip in the power just were they should 
be.}, where they represent about a 40\% diminuation of the power at their 
centers.  The ``notches'' do appear to be just visible in a sensitive 430-MHz 
observation of Blaskiewicz \etal\ (1991: see their Fig.~22), though they are 
not seen at 1.4 GHz either in this paper or in Rankin \& Rathnasree.  This 
may be due to the overall weakening of the broad low-level emission feature 
that follows the MP as well as the poorer signal-to-noise of the 21-cm 
observations.  Neither, unfortunately, were the recent observations of 
Weisberg \etal\ (2003) sufficiently sensitive to detect the features.  

\begin{figure}
\begin{center}
\epsfig{file=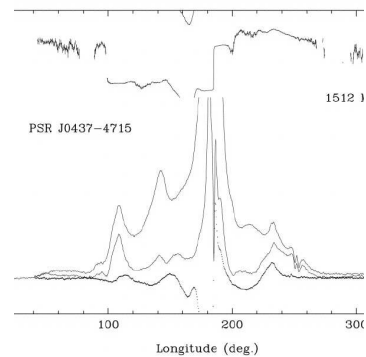, height=6.0cm}
\caption{Polarized pulse profile for J0437$-$4715 at 1512~MHz from 
Navarro \etal\ (1997); their Fig.~4.
\label{fig:0437a}}
\end{center}
\end{figure}

\begin{figure}
\begin{center}
\epsfig{file=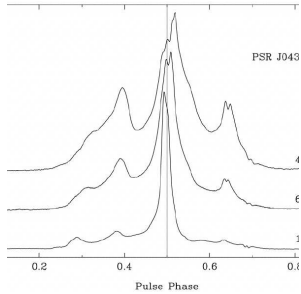, height=5.0cm}
\caption{Total-intensity profiles for J0437$-$4715 at 438, 660 and 
1512~MHz, which have been aligned according to the zero-crossing 
point of the anti-symmetric circular polarization under their central 
components from Navarro \etal\ (1997); their Fig.~5.  Note that the 
``notches'' almost align---and that they could even do so more closely 
if only a slightly different alignment had been used.  
\label{fig:0437b}}
\end{center}
\end{figure}

Similar features have also been observed by Navarro \etal\ (1997) in the
average profile of the millisecond pulsar J0437$-$4715.  These 
``notches'' fall on the trailing edge of this star's very broad profile, some 
70$\deg$ after the bright central component.  Here, the pair is separated 
by about 3.3$\deg$, the half-power widths are about a degree, and the 
fractional intensity decrease at their centers is about 50\%.  These 
characteristics are clearly seen in the 1512-MHz polarization profile of 
Navarro \etal\ (their Fig.~4) which we reproduce here as Fig.~\ref{fig:0437a}. 
However, this paper reports additional detections at frequencies of 438 
and 660~MHz, and the three total-power profiles appear together in our 
Fig.~\ref{fig:0437b}, aligned (according to their Fig.~5 caption) on the 
basis of the anti-symmetric circularly polarized signature associated with 
the central component.  One can see that the ``notches'' at the three 
frequencies do not quite align, falling slightly earlier at 660 and 1512 MHz.  
However, were the profiles aligned more nearly according the the centroids 
of the central component, the ``notches'' might well fall at just the same 
profile phase.  It would be interesting to see the results of a timing alignment 
as carried out for B0950+08 above.  Finally, the features do appear to 
evolve somewhat with frequency, becoming wider and more separated 
at lower frequencies. 

It is noteworthy that, for all three pulsars, these unprecedented features 
have roughly the same angular width, implying that their overall scale is 
some nearly fixed fraction of the rotational cycle of the star [as, apparently, 
is also the case for pulsar microstructure (\eg, Popov \etal\ 2002)]. Furthermore, 
for the two pulsars in which the features have been detected at more than 
one frequency, they seem to appear at exactly the same longitude.  Again, 
this is a well-known feature of pulsar microstructure, implying that the two 
phenomena derive from the same physical regions in the pulsar magnetosphere. 
Pulsar microstructure generally shows the same width regardless of frequency, 
however, there is evidence for the width of the notch features decreasing 
with frequency.

\medskip

B0950+08 and B1929+10 have remarkably similar properties. Both 
are very bright, have quarter-second rotational periods, and are very nearby.
Both pulsars have detectable emission over an unusually broad fraction of 
their rotation cycle and both exhibit interpulses.  With a spin period of 5.75~ms, 
J0437$-$4715 has very different spin-down properties; however, it is also 
nearby, extremely bright, and, while it has no interpulse, it has detectable emission 
over more than $180^\circ$ of longitude. 

The identification of these features in the three brightest pulsars with broad 
pulse profiles suggests that such features may be common among pulsars 
with broad profiles. As they are apparently seen over a substantial portion 
of the radio band and evolve little with frequency, it would seem that they 
cannot be attributed to absorption in any usual sense.  Thus it would seem 
that they are either due to some very puzzling aspect of the emission process 
or represent an equally puzzling obscuration of the emission along our sight 
line. While we cannot more than speculate about the causes of these 
features, a very novel mechanism is being suggested by Wright (2003) in a 
companion paper.  Clearly, few pulsars show emission so far from the main 
pulse or interpulse, so observing this feature in more pulsars will be very
difficult until the advent of more sensitive pulsar instruments such as the SKA. 
However, future more sensitive multi-frequency observations of the pulsars 
discussed above, in addition to polarimetric observations, may yield more
clues as to the origin of these unusual features.  

\bigskip

\section*{Acknowledgments}

We thank Leszek Nowakowski and Geoff Wright for discussions and critical 
comments on the manuscript.  One of us (JMR) wishes to acknowledge 
support from US National Science Foundation Grant AST 99-87654.  The 
Arecibo Observatory is operated by Cornell University under contract to the 
US NSF.  

{}


\begin{thebibliography}{}
\bibitem[Backer 1976]{} Backer, D. C. 1976, ApJ, 209, 895
\bibitem[]{} Blaskiewicz, M., Cordes, J.~M., \& Wasserman, I. 1991, ApJ, 370, 643
\bibitem[Brisken \etal\ 2000]{brisk2000} Brisken, W.~F., Benson, J.~M., Beasley, A.~J., 
	Fomalont, E.~B., Goss, W.~M., \& Thorsett, S.~E. 2000, ApJ, 541, 959
\bibitem[Brisken \etal\ 2002]{brisk2002} Brisken, W.~F., Benson, J.~M., Goss, W.~M., 
 	Thorsett, S.~E. 2002, ApJ, 571, 906  
\bibitem[]{} Gould, D.~M., \& Lyne, A.~G. 1998, MNRAS, 301, 235.  	\bibitem[]{} Hankins, T.~H., \& Rankin, J.~M. 2003, ApJ, preprint  
\bibitem[Hobbs 2002]{hobbs2002} Hobbs, G. PhD Thesis. 2002. University of Manchester 
\bibitem[]{} De Luca, A., Mignani, R.~P., Caraveo, P.~A., 2003, "Radio Pulsars" 
	ASP Conference Series, eds.,  M. Bailes, D.~J. Nice, S.~E. Thorsett, 302, 359
\bibitem[Lyne \& Manchester 1988]{lyne1988} Lyne, A.~G.~\& Manchester, R.~N.\ 1988, 
	MNRAS, 234, 477
\bibitem[]{} Mignani, R.~P., De Luca, A., Caraveo, P.~A., \
	\& Becker, W. 2002, ApJ, 580, L147  
\bibitem[Navarro \etal\ 1997]{navarro97} Navarro, J., Manchester, R.~N., 
	Sandhu, J.~S., Kulkarni, S.~R., \& Bailes, M.\ 1997, ApJ, 486, 1019
\bibitem[]{} Nowakowski, L.~A., Bhat, N.~D.~R., \& Lorimer, D.~R. 2003, 
	Arecibo Observatory Newsletter \#36  
\bibitem[]{} Phillips, J.~A. 1990, ApJ, 361, L57  
\bibitem[Popov \etal\ (2002)]{} Popov, M. V., Bartel, N., Cannon, W. H., Novikov, A. Yu., 
	Kondratiev, V. I., \& Altunin, V. I. 2002, A\&A, 196, 171
\bibitem[Radhakrishnan \& Cooke 1969]{rad1969} Radhakrishnan, V.~\& Cooke, D.~J.\ 1969, 
	Ap. Lett., 3, 225
\bibitem[Rankin 1983]{rankin1983} Rankin, J.~M.\ 1983, ApJ, 274, 333
\bibitem[Rankin 1993]{rankin1993} Rankin, J.~M.\ 1993, ApJ, 405, 285 
\bibitem[Rankin \& Rathnasree 1997]{rankin97} Rankin, J.~M.~\& Rathnasree, N.\ 1997, 
	J. Astrop. \& Astr., 18, 91
\bibitem[Ruderman \& Sutherland 1975]{ruderman1975} Ruderman, M.~A.~\& 
	Sutherland, P.~G. 1975, ApJ, 196, 51 
\bibitem[Sturrock 1971]{sturrock1971} Sturrock, P.~A.\ 1971, ApJ, 164, 529
\bibitem[Taylor \etal\ 1993]{ppcat} Taylor, J.~H., Manchester, R.~N., \& Lyne, A.~G. 1993, 
	ApJ Suppl., 88, 529
\bibitem[Taylor \& Weisberg 1989]{tay89} Taylor, J.~H. \& Weisberg, J.~M. 1989, ApJ, 345, 132 
\bibitem[Weisberg \etal\ 1999]{} Weisberg, J.~M., Cordes, J.~M., Lundgren, S.~C., 
	Dawson, B.~R., Despotes, J.~T., Morgan, J.~J., Weitz, K.~A., Zink, E.~C., \& 
	Backer, D.~C.1999, ApJ Suppl., 121, 171  
\bibitem[Weisberg \etal\ 2003]{} Weisberg, J.~M., Cordes, J,~M., Kuan, B., Devine, K. E., 
	Green, J.~T., \& Backer, D.~C. 2003, ApJ, in press  
\bibitem[]{} Wang, F.~Y.-H., \& Halpern, J.~P. 1997, ApJ, 482, L159
\bibitem[]{} Wright, G. A. E. 2003, MNRAS, submitted, astro-ph/0311467
\bibitem[]{} Zavlin, V.~E., Pavlov, G.~G., Sanwal, D., Manchester, R.~N., Truemper, J.,
	Halpern, J.~P., Becker, W., 2002, ApJ, 569, 894
\end{thebibliography}
\end{document}